\documentclass[twocolumn,showpacs,preprintnumbers,amsmath,amssymb,prl]{revtex4}

\usepackage[dvips]{graphicx}
\usepackage{dcolumn}
\usepackage{amssymb}
\usepackage{amsfonts}
\usepackage{bm}

\begin{document}

\preprint{APS}
\title{Effect of motion of the scatterers  on localization: quasi localization and quasi mobility edge}
\author{E. Kogan}
\email{kogan@mail.biu.ac.il}
\affiliation{Jack and Pearl Resnick
Institute, Physics Department, Bar Ilan University, Ramat Gan 52900,
Israel}
\date{\today}

\begin{abstract}
We  study kinetics of electrons,   scattered by heavy particles
undergoing slow diffusive motion. In a three-dimensional space we
claim the existence of the crossover region (on the energy axis),
which separates the states with  fast diffusion and the states with
slow diffusion; the latter is determined by the dephasing time. In a
two-dimensional space the diffusion coefficient for any value of
energy is determined by the dephasing time.
\end{abstract}
\maketitle

Consider  electrons in an inhomogeneous media.  When discussing the
kinetics of the electrons, the quantity, we usually start from in
the theoretical description, is the transport relaxation time
$\tau$, calculated in Born approximation. Using Boltzmann equation
we can obtain the relation between this relaxation time and the
diffusion coefficient
\begin{eqnarray}
\label{diffusion}
 D_0=\frac{1}{3}v^2\tau,
\end{eqnarray}
where  $v$ is the electron velocity. However, the  Boltzmann
equation is valid provided $ E\tau \gg 1$.
If we take into account that $1/\tau(E)$ typically
decreases slower than the first power of energy when the latter goes
to zero, we see that however weak is the scattering, the condition
of the applicability of Boltzmann equation is broken near the band bottom. It is well known since
the seminal works of N. F. Mott \cite{mott}, about the existence of
the mobility edge $E_c$, that is the energy which separates the
states with finite diffusion coefficient and states with the
diffusion coefficient being exactly equal to zero. All this is true
provided the disorder is static. Natural question arises: what
happens with this picture when the scatterers slowly move.

To answer this question  we need some theory of localization. As
such we'll use the self-consistent localization theory by Vollhard
and W\"olfle \cite{vollhardt}. Of crucial importance in the above
mentioned theory are maximally crossed diagrams (the sum of all such
diagrams is called Cooperon) for the two-particle Green function.
The calculations of these diagrams for the case of moving scatterers
were done in the paper by Golubentsev \cite{golubentsev}. So in the
first part of the present paper we reproduce the results by
Golubentsev (plus some additional interpretation). In the second
part we use the results for the Cooperon as an input for the
self-consistent localization theory, which we  modify to take into
account the slow motion of scatterers. In the third part we discus
the results obtained.

The electrons are scattered by the potential
\begin{eqnarray}
V(r,t)=V\sum_a\delta\left(r-r_a(t)\right).
\end{eqnarray}
Define the correlator
\begin{eqnarray}
K(r-r',t-t')=<V(r,t)V(r',t')>.
\end{eqnarray}
In the leading approximation in the scatterers density we have for
the Fourier component of the correlator
\begin{eqnarray}
K(q,t)=V^2\left\langle\int\exp\left\{ (i{\bf q}({\bf r}-{\bf r}') \right\}\right.\nonumber\\
\left. \times  drdr'\sum_a\delta\left({\bf r}-{\bf r}_a(t)\right)
\sum_{a'}\delta\left({\bf r}'-{\bf r}_{a'}'(0)\right)\right\rangle\nonumber\\
=V^2\sum_a\left\langle\exp\left\{ (iq({\bf r}_a(t)-{\bf
r}_a(0))\right\}\right\rangle =nV^2f({\bf q},t),
\end{eqnarray}
where $n$ is the scatterer density. We consider the case when the
scatterers undergo slow diffusive motion. In the ballistic case
\begin{eqnarray}
f({\bf q},t)=\exp\left(-\frac{{\bf q}^2}{6}<{\bf
v}^2>t^2\right),\qquad |t|\ll\tau_{imp},
\end{eqnarray}
In the diffusive case
\begin{eqnarray}
f({\bf q},t)=\exp\left(-\frac{{\bf q}^2}{2}D_{imp}|t|\right),\qquad
|t|\gg\tau_{imp},
\end{eqnarray}
where $D_{imp}=<v^2>\tau_{imp}/3$, and $\tau_{imp}$ is the
scatterers free path time.

 For the Cooperon  we get \cite{golubentsev}
\begin{eqnarray}
\label{cooperon}
C_E({\bf q})= \int_0^{\infty}\exp\left\{-D(E)q^2t
-\frac{1}{\tau} \int_0^t(1-f_{t'})dt'\right\}dt,
\end{eqnarray}
where $E$ is the energy of each of  the two electron lines in
Cooperon diagram, and $q$ is the sum their momenta (see Fig. 1).
Also
\begin{eqnarray}
\frac{1}{\tau}=nV^2\frac{k^2}{\pi v},
\end{eqnarray}
and
\begin{eqnarray}
f_t=\int\frac{d{\bf s}'}{4\pi}f(k_0(s-s'),t)\nonumber\\
 =\left\{\begin{array}{ll}y\left(\frac{t^2}{\tau_{\lambda}^2}\right) & |t|\ll\tau_{imp} \\
                         y\left(\frac{|t|\tau_{imp}}{\tau_{\lambda}^2}\right) & |t|\gg\tau_{imp}\end{array}\right.
\end{eqnarray}
where
\begin{eqnarray}
y(x)=\frac{1-e^{-x}}{x},\;
\tau_{\lambda}=\left(\frac{2}{3}k^2<v_{imp}^2>\right)^{-1/2}.
\end{eqnarray}

Eq. (\ref{cooperon}) can be easily understood if we compare diagrams
for the Diffuson (the sum of all ladder diagrams) and the Cooperon
on Fig. 1. The Diffuson does not have any mass because of Ward
identity. In the case of the Cooperon, the Ward identity is broken,
and the difference $[1-f(t)]$ shows how strongly. The interaction
line which dresses single particle propagator is given by static
correlator, and interaction line which connects two different
propagators in a ladder is given by dynamic correlator.
\begin{figure}
\includegraphics[angle=0,width=0.45\textwidth]{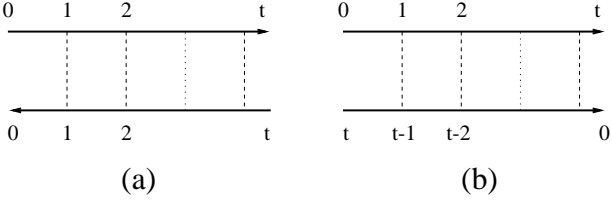}
\caption{Diagrams for the Diffuson (a) and the Cooperon (b). Solid
line is dressed electron propagator, dashed line connecting points
${\bf r},t$ and ${\bf r}',t'$ corresponds to $K({\bf r-r'},t-t')$.}
\end{figure}
The time-reversal invariance in the system we are considering is
broken due to dephasing; the diffusion pole of the particle-particle
propagator disappears, although particle-hole propagator still has a
diffusion pole, which is guaranteed by particle number conservation.

 In extreme cases, from Eq. (\ref{cooperon}) we obtain
\begin{eqnarray}
\label{cooperon2}
C_E({\bf q})=
\int_0^{\infty}\exp\left[-D(E){\bf q}^2t-t^3/\tau_{\varphi}^3(E)\right]dt,\\
\text{at}\qquad\tau_{\lambda}^2\tau\ll\tau_{imp}^3,\;\tau\ll\tau_{\lambda}\nonumber
\end{eqnarray}
\begin{eqnarray}
\label{cooperon22}
 C_E({\bf q})=
\int_0^{\infty}\exp\left[-D(E){\bf q}^2t-t^2/\tau_{\varphi}^2(E)\right]dt,\\
\text{at}\qquad\tau_{\lambda}^2\tau\gg\tau_{imp}^3,\;\tau\tau_{imp}\ll\tau_{\lambda}^2\nonumber
\end{eqnarray}
where in the ballistic case
\begin{equation}
\label{g}
\tau_{\varphi}=(3\tau\tau_{\lambda}^2)^{1/3}\qquad
\text{at}\qquad \tau_{\lambda}^2\tau\ll\tau_{imp}^3,
\end{equation}
and in the diffusive case
\begin{equation}
\label{g2}
\tau_{\varphi}=(2\tau\tau_{\lambda}^2\tau_{imp}^{-1})^{1/2}\qquad
\text{at}\qquad \tau_{\lambda}^2\tau\gg\tau_{imp}^3.
\end{equation}
Thus we obtain the crucial parameter  - the dephasing time
$\tau_{\varphi}$.

The results  for the dephasing time (up to a numerical factors of
order of one) can be understood using simple qualitative arguments.
Consider ballistic regime. If a single collision leads to the
electron energy change $\delta E$, the dephasing time could be
obtained using relation \cite{altshuller}
\begin{equation}
\tau_{\varphi}\delta E\sqrt{\frac{\tau_{\varphi}}{\tau}}\sim 2\pi,
\end{equation} where $\tau_{\varphi}/\tau$ is just the
number of scattering acts during the time  ${\tau_{\varphi}}$. So in
this case
\begin{equation}
\label{toy} \frac{1}{\tau_{\varphi}^3}\sim\frac{(\delta E)^2}{\tau}.
\end{equation}
If we notice that $1/\tau_{\lambda}$ is the averaged electron energy
change in a single scattering act $\delta E$, we immediately regain
Eq. (\ref{g}).

Inserting Eq. (\ref{cooperon2}) into the self-consistent equation,
for the diffusion coefficient $D$  we obtain equation
\begin{eqnarray}
\label{ret}
 \frac{D_0(E)}{D(E)}=1+\frac{1}{4\pi^2 mk} \sum_{\bf
q}C_E({\bf q})
\end{eqnarray}
where $D_0$ is the diffusion coefficient calculated in Born
approximation (Eq. (\ref{diffusion}) and the momentum cut-off $|{\bf
q}|<1/\ell$  is implied, where $l=k\tau/m$ is the  mean free path.
 Thus we obtain
\begin{eqnarray}
\label{re}
\frac{D_0}{D}=1+\frac{1}{\pi mk}
 \int_0^{\infty}dt
\int_0^{1/l}dq\; q^2\;\exp\left[-Dq^2t -g(t) \right],\nonumber\\
\end{eqnarray}
where
\begin{eqnarray}
g(t)=\frac{1}{\tau}\int_0^t(1-f_{t'})dt'.
\end{eqnarray}
(In the particular case of ballistic regime $g(t)=t^3/\tau_{\varphi}^3$, and in the
diffusive regime $g(t)=t^2/\tau_{\varphi}^2$.)
The  Eq. (\ref{re}) can be
presented as
\begin{eqnarray}
\label{ret2}
\frac{D_0}{D}=1+X_{IR}\frac{D_0}{D}\nonumber\\
\cdot \int_0^{\infty}dx\int_0^{1}dy\;
y^2\;\exp\left[-xy^2-g\left(\frac{xl^2}{D}\right)\right],
\end{eqnarray}
where
\begin{equation}
X_{IR}=\frac{3}{4\pi E^2\tau^2}
\end{equation}
is the Iofe-Regel parameter.
Further on we'll consider Eq. (\ref{ret2}) for
\begin{eqnarray}
\label{ggt}
\tau_{\varphi}\gg\tau.
\end{eqnarray}
To  solve  the equation we take into account that the function
is equal to zero for $t=0$ and assume that it becomes of the order of 1 for $x=D/D_{\varphi}$,
where we introduced
$ D_{\varphi}=l^2/\tau_{\varphi}$.
Notice, that the condition (\ref{ggt}) can be presented as
$D_{\varphi}\ll D_0$.

Let us analyze the behavior of the r.h.s. of Eq. (\ref{ret2}) as a function of $D$.
For $D\gg D_{\varphi}$ the second term in the exponent in Eq. (\ref{ret2}) (which represents dephasing) becomes irrelevant, and we get
\begin{eqnarray}
\label{asim}
\text{r.h.s.}(\ref{ret2})=1+X_{IR}\frac{D_0}{D}.
\end{eqnarray}
Substituting this result into Eq. (\ref{ret2}) we obtain the solution
\begin{equation}
\label{m}
 D=D_0(1-X_{IR}).
\end{equation}
This solution is valid, provided that  $X_{IR}<1$, that is   $E>E_c$, where the mobility edge $E_c$ is obtained from the equation
\cite{vollhardt}
\begin{equation}
\label{mit2}
E_c\tau(E_c)=\sqrt{3/4\pi},
\end{equation}
which is just the Iofe-Regel criterium
For $X_{IR}>1$, thinking in terms of graphical method of solution, we see that
the curve,  representing the asymptotic formula (\ref{asim}), is above the curve, representing the l.h.s. of Eq. (\ref{ret2}), and they never cross. However, Eq. (\ref{asim}) is no longer valid for $D\leq D_{\varphi}$. In particular, when $D\ll D_{\varphi}$, it is the first term in the exponent which is irrelevant, and we get
\begin{eqnarray}
\label{bm}
\text{r.h.s.}\sim \frac{D_0}{D_{\varphi}}
\end{eqnarray}
This equation guarantees the existence of solution for  $X_{IR}>1$. Because the only scale parameter in the r.h.s. is the quantity $D/D_{\varphi}$, the deviations from the asymptotic formula (\ref{asim}) appear only for $D\leq D_{\varphi}$, and the solution for $X_{IR}>1$ is
\begin{eqnarray}
\label{sim}
D= D_{\varphi}/\alpha,
\end{eqnarray}
where $\alpha$ is the solution of equation
\begin{eqnarray}
\label{ret222}
1= X_{IR}\int_0^{\infty}dx\int_0^{1}dy\;
y^2\;\exp\left[-xy^2-g\left(\alpha x\tau_{\varphi}\right)\right].
\end{eqnarray}
In particular, deep in the "dielectric region" ($E\tau\ll1$),
the solution of the self-consistent equations is
\begin{equation}
\label{diel}
 D=\frac{4\pi}{\int_0^{\infty}\exp [-g(z\tau_{\varphi})]dz} E^2\tau^2 D_{\varphi}.
\end{equation}
Eqs. (\ref{m}) and (\ref{sim}) cover the
whole region of change of the Iofe-Regel parameter save the narrow
cross-over region near $X_{IR}\approx 1$.

The influence of static disorder in the spaces with dimensionality 2
and 1 is drastically different from that in the space with the
dimensionality 3, considered above. In one- and two-dimensional
spaces all the states are localized cite{efetov} (provided the
disorder is static). Again we ask ourselves, what happens with this
picture when the scatterers undergo slow diffusive motion. According
to the self-consistent localization theory \cite{vollhardt}, for
the space of dimensionality $d$ the Eq. (\ref{ret2}) becomes
\begin{eqnarray}
\label{ret22}
\frac{D_0}{D}=1+X_{IR}\frac{D_0}{D}\nonumber\\
\cdot \int_0^{\infty}dx\int_0^{1}dy\;
y^{d-1}\;\exp\left[-xy^2-g\left(\frac{xl^2}{D}\right)\right],
\end{eqnarray}
and the Eq. (\ref{diffusion}) is
$ D_0=\frac{1}{d}v^2\tau$.
Analyzing behavior of the r.h.s., for $D\gg D_{\varphi}$ we obtain
\begin{eqnarray}
\label{asim2}
\text{r.h.s.}(\ref{ret22})=1+X_{IR}\frac{D_0}{D}\ln\left(\frac{D}{D_{\varphi}}\right).
\end{eqnarray}
Again thinking in terms of graphical method of solution, we see that
the curve,  representing the asymptotic formula (\ref{asim2}), is for any $X_{IR}$ above the curve, representing the l.h.s. of Eq. (\ref{ret22}), unless $D\sim D_{\varphi}$.
Thus the solution for any value of the Iofe-Regel parameter is
\begin{eqnarray}
D=\beta D_{\varphi},
\end{eqnarray}
where $\beta$ is the solution of the equation
\begin{eqnarray}
\label{ret333}
1= X_{IR}\int_0^{\infty}dx\int_0^{1}dy\;
y\;\exp\left[-xy^2-g\left(\beta x\tau_{\varphi}\right)\right].
\end{eqnarray}

\section{Conclusions}

We considered the influence of slow diffusive motion of scatterers
on the localization of electrons.
In this case, like in the case of purely elastic scattering, the
diffusion coefficient drastically differs for the energies below and above the mobility edge,
the latter being found from the Iofe-Regel criterium. Above the mobility edge we have fast diffusion, and the defasing is irrelevant. Below the mobility edge the diffusion coefficient is inversly proportional to the diffusion time.

Now we would like to mention some
possible generalization and applications of these ideas. First, they
are completely applicable for the case of ballistic motion of
scatterers, say to gas, consisting of heavy classical particles and
electrons. The results are identical to those obtained above in the
ballistic regime. Second, in the previous publication \cite{kog}, we
studied the influence of dephasing on the Anderson localization of
the electrons in magnetic semiconductors, driven by spin
fluctuations of magnetic ions. There the role of heavy particles was
played by magnons; complete spin polarization of conduction
electrons prevented magnon emission or absorption processes, and
only the processes of electron-magnon scattering being allowed.
Finally, the results obtained can be applied for studying kinetics
of classical waves, especially light. This last application is
particularly appealing, taking into account that the results of
Golubentsev \cite{golubentsev} were obtained for light waves.

\end{document}